\documentclass[aps,prl,reprint,twocolumn]{revtex4-2}

\usepackage{amsmath}
\usepackage{amsfonts}
\usepackage{amssymb}
\usepackage[usenames,dvipsnames]{color}
\usepackage[colorlinks=true,linkcolor=Maroon,citecolor=Maroon,urlcolor=Maroon]{hyperref}
\usepackage{graphicx}
\usepackage{xcolor}
\usepackage{comment}

\usepackage{enumitem}
\usepackage{amsthm}          
\usepackage{tikz,pgfplots}

\theoremstyle{remark}             

\makeatletter
\def\ps@pprintTitle{%
 \let\@oddhead\@empty 
 \let\@evenhead\@empty
 \def\@oddfoot{}%
 \let\@evenfoot\@oddfoot}
 \makeatother 

\newcommand{\A}{\mathbf{A}}

\newcommand{\R}{\mathbf{R}}

\newcommand{\PP}{\mathbf{P}}

\newcommand{\I}{\mathbf{I}}
\newcommand{\Q}{\mathbf{Q}}
\newcommand{\U}{\mathbf{U}}
\newcommand{\V}{\mathbf{V}}

\newcommand{\x}{\mathbf{x}}

\newcommand{\vv}{\mathbf{\hat{v}}}
\newcommand{\uu}{\mathbf{\hat{u}}}

\newcommand{\et}{\boldsymbol{\eta}}
\newcommand{\PPi}{\boldsymbol{\Pi}}
\newcommand{\Sig}{\boldsymbol{\Sigma}}
\newcommand{\Lamb}{\boldsymbol{\Lambda}}

\begin{document}

    \title{Eigenvector Geometry as a New Route to Criticality \\ in Random Multiplicative Systems}

    \author{Virgile Troude}
    \author{Didier Sornette}
    \affiliation{Institute of Risk Analysis, Prediction and Management (Risks-X),
    Academy for Advanced Interdisciplinary Sciences,
    Southern University of Science and Technology, Shenzhen, China}

\begin{abstract}
    Heavy-tailed fluctuations and power law distributions pervade physics, biology, and the social sciences, 
    with numerous mechanisms proposed for their emergence. 
    Kesten processes, which are multiplicative stochastic recursions with additive noise or reinjection, provide a canonical explanation, 
    where power law tails arise from transient supercritical excursions as eigenvalues intermittently cross the stability boundary. 
    Here we uncover a distinct and more general mechanism in multidimensional systems: \emph{non-normal eigenvector amplification}. 
    In random non-normal matrices, the non-orthogonality of eigenvectors, 
    quantified at each time step by the condition number $\kappa_t$ in Kesten-like processes, induces transient growth that 
    increases the effective Lyapunov exponent $\gamma \to \gamma + \mathbb{E}\left[\ln \kappa_t \right]$ 
    and lowers the tail exponent $\alpha \simeq -2\gamma / \sigma_{\kappa}^2$, where 
    $\mathbb{E}\left[\ln \kappa_t \right]$ and $\sigma_{\kappa}^2$ are respectively the mean and variance of $\ln \kappa_t$. 
    As the system dimension $N$ grows, $\kappa$ typically increases proportionally, making non-normal amplification the dominant source of scale-free behavior. 
    We illustrate this mechanism in polymer stretching in turbulent flows,
    where intermittent extensions arise from eigenvector amplification of velocity gradients.    
\end{abstract}

\maketitle

Heavy-tailed distributions are a hallmark of complex systems, arising in physics, finance, economics, biology, and many other domains.  
A canonical mechanism for their emergence is provided by the class of stochastic recursions studied by Kesten~\cite{kesten1973random}, where the interplay between multiplicative growth and additive noise naturally produces stationary distributions with power law tails.  
Such \emph{Kesten processes} have since become a cornerstone in the theoretical understanding of scale-free phenomena, with broad applications ranging from physics \cite{deCalan1985,Takayasu1997,BlankSolomon2000}, biology \cite{Statman2014}, to financial markets~\cite{LuxSornette2002,lux2006financial,Gabaix2009,Derman2010}, and to models of wealth and income distribution~\cite{Sornette1998,benhabib2016wealth}.

The classical understanding attributes the emergence of power laws in Kesten-type processes to \emph{spectral supercriticality}, i.e., episodes in which eigenvalues of the random multiplicative operator transiently cross the unit circle.  
In this view, power law tails reflect rare bursts of exponential growth balanced by global stability on average.  

In this paper, we uncover a complementary and more general mechanism in multidimensional processes: \emph{non-normal eigenvector amplification}.  
When random matrices are non-normal (not unitarily diagonalizable), transient growth can occur even when all eigenvalues lie strictly within the unit circle.  
This effect, arising from the non-orthogonality of eigenvectors, provides a distinct and generic route to heavy-tailed stationary distributions.  
Our analysis shows that non-normality not only lowers the tail exponent, thereby producing fatter tails, but can also shift the effective Lyapunov exponent,
destabilizing the system in a way fundamentally different from spectral criticality,
leading in some limit to true instability (criticality).

The power law mechanism rooted in the generic geometry of eigenvectors in high-dimensional random systems provides a unified explanation for heavy-tailed distributions in systems governed by multiplicative interactions, from turbulent flows to wealth and income dynamics. By identifying non-normal amplification as a fundamental source of heavy tails, our work extends the theoretical scope of Kesten processes and clarifies their relevance to the statistics of complex systems.

Our work builds on a broad literature that first recognized the dynamical significance of non-normal operators. 
Pioneering studies in hydrodynamic stability and atmospheric dynamics by Farrell and Ioannou~\cite{Farrell1988,Farrell1989,FarrellIoannou1996a,FarrellIoannou1996b,FarrellIoannou2003,FarrellIoannou2007} 
and in fluid mechanics by Trefethen and collaborators~\cite{Trefethen1993,Embree2005} 
showed that non-normality can induce large transient amplifications even when all eigenvalues indicate linear stability. 
Related ideas were later extended to other fields, such as subcritical magnetic dynamos~\cite{fedotov2003nonnormal}. 
Here we generalize this perspective beyond hydrodynamics and linear response, showing that non-normal eigenvector amplification 
constitutes a universal route to heavy-tailed fluctuations in high-dimensional stochastic systems.


\vskip 0.2cm
\noindent
\it Kesten formalism.
We consider the $N$-dimensional Kesten process
\begin{equation}
    \x_{t+1} = \A_t \x_t + \et_t, 
    \label{fhwtbgq}
\end{equation}
where $\A_t$ is an i.i.d. sequence of random matrices and $\et_t$ is an additive noise term.
To apply Kesten formalism \cite{kesten1973random},
$\eta_t$ must be independent of the multiplicative matrices $A_t$,
and must not vanish identically,
i.e. $\mathbb{P}(\eta_t=0)<1$,
so that re-injection occurs with nonzero probability.
Moreover, one of our results is that power law tails emerge even far from spectral criticality for all $\A_t$'s.
Beyond these conditions, the detailed distribution of $\eta_t$
(e.g., whether it has finite mean or finite variance)
does not affect the tail exponent of the stationary distribution,
which is determined by the multiplicative dynamics.
This remains true provided that the distribution of $\eta_t$ is not heavier-tailed than
the distribution resulting from the multiplicative mechanism itself.
In particular, the results are unchanged if $\eta_t=\eta\neq 0$ is a fixed nonzero constant vector.

The dynamic long-term behavior is controlled by
\begin{equation}
    \pi_t = \|\PPi_t\| \quad  {\rm with}
    \quad \PPi_t = \A_t\PPi_{t-1}~,
    \label{fhbgvqa}
\end{equation}
where $\Pi_t$ denotes the product of matrices up to time $t$ and $\pi_t$ is its $L_2$-norm.
The stability of the process is characterized by the \emph{Lyapunov exponent}~$\gamma$, defined as
\begin{equation}
    \gamma := \lim_{t \to \infty} \frac{1}{t}\,\mathbb{E}\!\left[\ln \pi_t\right] .
    \label{fdhtgbq}
\end{equation}
A negative exponent ($\gamma<0$) implies asymptotic stability,
with convergence to a stationary distribution, whereas a positive exponent ($\gamma>0$) indicates instability,
with the system diverging and growing exponentially in time.

The \emph{tail exponent} $\alpha$ quantifies the heaviness of the tail of the stationary distribution, namely
\begin{equation}
    \mathbb{P}[\mathbf{n}\cdot \mathbf{x}_t > x_n] \sim x_n^{-\alpha}, \quad x_n \to \infty ~,
    \label{hy3ujun3}
\end{equation}
where $\mathbf{n}\cdot \mathbf{x}_t$ denotes any projection (the same asymptotics hold for the $L_2$-norm).
Consider the function 
\begin{equation} \label{eq:phi_alpha}
    \phi(\alpha) := \lim_{t\to\infty} \frac{1}{t}\,\ln \mathbb{E}\!\left[\pi_t^\alpha\right]  ~,
\end{equation}
which is convex, satisfies $\phi(0)=0$, and has derivative $\phi'(0)=\gamma$.  
For $\gamma<0$, the convexity of $ \phi(\alpha)$ ensures that there exists a unique $\alpha>0$ solving $\phi(\alpha)=0$,
which determines the tail exponent of the stationary distribution \cite{FurstenbergKesten1960,kesten1973random}.
For $\gamma \geq 0$, no positive solution exists and the process fails to admit a stationary heavy-tailed regime.


\vskip 0.2cm
\noindent
{\it Normal Kesten processes: spectral criticality}.
    When the random matrices $\A_t$ are \emph{normal} (i.e. unitarily diagonalizable), they can be written as
    $\A_t = \U_t \Lamb_t \U_t^\dagger$ with $\U_t$ unitary and $\Lamb_t$ diagonal.
    Then the product norm has an upper bound given by
   $\pi_t  \leq  \prod_{s=1}^t\|\A_s\| = \prod_{s=1}^t~\rho_s$,
    since, for normal matrices, the $L_2$-norm $\|\A_t\|$ of $\A_t$ is given by its spectral radius $\rho_t$.
    One can note that $\prod_{s=1}^t~\rho_s = e^{R_t}$, where $R_t= \sum_{s=1}^t\ln\rho_t$,
    and by the \textit{Central Limit Theorem} (CLT), we have $R_t\sim\mathcal{N}(t\ln\rho,t\sigma_\rho^2)$,
    where $\ln\rho$ and $\sigma_\rho ^2$ are respectively the expected value and variance of the logarithm of the spectral radius.
    Using the monotonicity of the logarithm and the convexity of $\phi(\alpha)$  \eqref{eq:phi_alpha},
    we obtain the following bounds for the Lyapunov and tail exponents
    \begin{equation}
        \gamma \leq   \ln\rho , \qquad
        \alpha \geq  -\frac{2\ln\rho}{\sigma_\rho^2}~~{\rm for}~\ln\rho <0~.
        \label{fhwrbg}
    \end{equation}

    The upper bound for $\gamma$ and the lower bound for $\alpha$ are reached for the one-dimension case for which 
    the Kesten process (\ref{fhwtbgq}) reduces to
    $x_{t+1} = \rho_t x_t + \eta_t$. When the multiplicative noise is
    lognormally distributed, $\ln \rho_t \sim \mathcal{N}(\ln \rho, \sigma_\rho^2)$,
    expression (\ref{fdhtgbq}) gives $\gamma = \ln\rho$ and the solution of the equation $\phi(\alpha)=0$ is exactly
    \begin{equation}
        \alpha = -\frac{2\ln\rho}{\sigma_\rho^2}~, ~~~{\rm for}~\ln\rho <0~.
        \label{fhrgbq}
    \end{equation}
    Hence, the one-dimensional Kesten process represents the \emph{worst-case scenario} of the multidimensional 
    Kesten process with normal matrices since the bounds (\ref{fhwrbg}) are exactly attained. 
    Intuitively, the power law tail originates from the fact that sequences 
    of successive $\rho_t>1$, which generate transient exponential growth, have an exponentially small probability to occur
  as a function of their durations, so that the stationary distribution reflects a balance between rare but strong amplification bursts 
    and global stability enforced by $\ln \rho < 0$~\cite{Sornette2006}.
    In other words, the heavy tails arise from eigenvalues stochastically crossing the unit circle,
    placing the system for brief periods in a spectrally supercritical state.
  
  \vskip 0.2cm
   \noindent
{\it Non-normal Kesten processes: eigenvector amplification}.
    When the random matrices $\A_t$ are \emph{non-normal}, i.e. not unitarily diagonalizable, an additional
    amplification mechanism appears.  
    Writing $\A_t = \PP_t \Lamb_t \PP_t^{-1}$,
    where $\Lamb_t$ is a diagonal matrix,
    the matrix $L_2$-norm of $\A_t$ is no more given by its spectral radius $\rho_t$,
    but depends on the condition number $\kappa_t = \|\PP_t\| \, \|\PP_t^{-1}\|$
    of the eigenbasis transformation, which quantifies the non-orthogonality of eigenvectors and the degree of non-normality
    ($\kappa_t>1$ for non-normal matrices and $\kappa_t=1$ for normal matrices).
    
    By the classical bound for diagonalizable matrices (see, e.g.,~\cite{HornJohnson2013}), with equality (and $\kappa=1$) when the eigenbasis is orthogonal, one has
    $\|\A_t\| \leq \kappa_t \,\rho_t $,.
    The $L_2$-norm $ \pi_t$ (\ref{fhbgvqa}) of the product $\PPi_t$ of matrices up to time $t$ is thus bounded as
    $\pi_t \leq \left(\prod_{s=1}^t\rho_s\right)\left(\prod_{s=1}^t\kappa_s\right)$.
   
    Similarly to the normal case, we can write $R_t = \sum_{s=1}^t\left(\ln\rho_t+\ln\kappa_t\right)$,
    and by the CLT, we have $R_t\sim\mathcal{N}\left(t\left(\ln\rho+\ln\kappa\right), t\sigma_0^2\right)$,
    where $\ln\rho$ and $\ln\kappa$ are respectively the expected value of $\ln\rho_t$ and $\ln\kappa_t$,
    and $\sigma_0^2 = \sigma_\rho^2 + \sigma_\kappa^2 + 2\text{Cov}\left[\ln\kappa_t,\ln\rho_t\right]$.
    Here, $\sigma_\rho^2$ and $\sigma_\kappa^2$ are respectively the variance of $\ln\rho_t$ and $\ln\kappa_t$
    and $\text{Cov}\left[\ln\kappa_t,\ln\rho_t\right]$ is the covariance between the log of the spectral radius,
    and the log of the degree of non-normality.
    This allows us to extend the previous bound \eqref{fhwrbg} to non-normal systems as
    \begin{equation}    \label{eq:lyapunov_bnd}
        \gamma \leq \ln\rho + \ln\kappa,\quad
        \alpha \geq -\frac{2}{\sigma_0^2}\left[\ln\rho+\ln\kappa\right],
    \end{equation}
    Thus, non-normality can (i) \emph{increase} the effective Lyapunov exponent, acting as a destabilizing force, and 
    (ii)  \emph{decrease} the tail exponent, thereby producing heavier power law tails.

\begin{figure}
    \centering
    \includegraphics[width=0.48\textwidth]{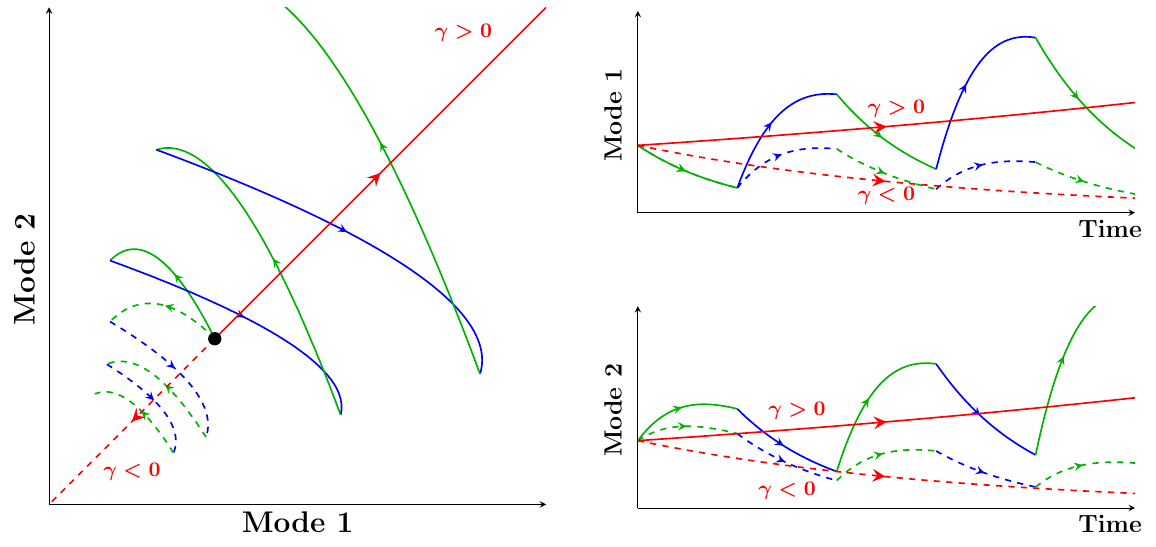}
    \caption{
        Schematic illustration of non-normal instability in a two-dimensional system, 
        with constant spectral radius $\rho$.  
        When the dominant mode alternates between the two axes,
        leading to cumulative reinjections and instability.
    }
    \label{fig:diagram}
\end{figure}

    Crucially, this mechanism operates \emph{even when the spectrum is strictly stable},
    providing a new and generic route to apparent criticality
    $\gamma \to 0$ and to power law distributions.  To see this,
    consider the case $\ln \rho < 0$ with $\sigma_\rho = 0$ and $\text{Cov}\left[\ln\kappa_t,\ln\rho_t\right]=0$ for normal matrices ($\kappa =1$): in this situation, 
    the bound \eqref{fhwrbg} and the exact expression \eqref{fhrgbq} for $\alpha$ diverge,
    indicating the absence of power law tails.  
    This is expected since the mechanism of intermittent supercriticality is absent.  
    When non-normality is present, stochastic fluctuations of the condition number,
    quantified by the mean $\ln\kappa$ and variance $\sigma_\kappa ^2$,
    increases $\gamma$ by $\ln\kappa$ and
    give rise to power law tails with effective exponent $\alpha \simeq -2\tfrac{\ln \rho + \ln\kappa}{\sigma_\kappa^2}$.
    The key intuition is that non-orthogonal eigenvectors allow transient amplifications:
    a vector multiplied by $\A_t$ can be stretched due to constructive interference between nearly aligned eigen-directions~\cite{Biancalani2017}.  
    In the product of random matrices, successive multiplications then act as random rotations,
    which can repeatedly project the state back into the most expanding direction.
    This recursive reinjection mechanism increases the effective Lyapunov exponent while simultaneously reducing the tail exponent,
    thereby amplifying large fluctuations, as illustrated in Figure~\ref{fig:diagram}.
    This simultaneous shift of the Lyapunov and tail exponents shows that non-normality can reshape both aspects of the dynamics at once, even when the spectrum itself remains strictly subcritical. 

\vskip 0.2cm
\noindent
{\it From Bounds to Explicit Approximations}.
We previously established an upper bound for the Lyapunov exponent and a corresponding lower bound for the tail exponent, and we showed that non-normality enlarges these bounds. We now go beyond bounding arguments and show that the increase in the Lyapunov exponent emerges directly within an explicit approximation, thereby indicating that the bounds are in fact tight. Specifically, by approximating the dynamics using the dominant contribution of each matrix in the product, we obtain an effective expression in which the impact of non-normality enters transparently. To proceed, we consider the $N$-dimensional case and introduce
the following decomposition
\begin{equation}    \label{eq:apx_dec_n_dim23}
    \A_t \;=\; \PP_t \,\Lamb_t\, \PP_t^\dagger~,
\end{equation}
where $\Lamb_t = \text{Diag}(\lambda_{i,t}\,|\,i=1,\dots,N)$ comprises the eigenvalues,
$\PP_t = \U_t \Sig_t \V_t^\dag$ is the eigenbasis transformation matrix,
$\Sig_t=\mathrm{Diag}(s_{1,t},\dots,s_{N,t})$ contains the singular values of $\PP_t$,
and $\U_t$ and $\V_t$ are unitary matrices.
The condition number quantifying the degree of non-normality is given by
$\kappa_t=s_{\max,t}/s_{\min,t}$, which is ratio of the largest to smallest singular values.

The dominant contribution to each matrix reads
\begin{equation}
    \A_t = \kappa_t\lambda^{\text{max}}_t \uu^{\text{max}}_{t}\left(\uu^{\text{min}}_t\right)^\dag + \cdots,
\end{equation}
where $\uu^{\text{max}}_{t}$ and $\uu^{\text{min}}_{t}$ are the singular column vectors
of $\U_t$ associated to the largest and smallest singular values of $\PP_t$,
and $\lambda^{\text{max}}_{t} = \vv^{\text{max}}_{t}\cdot\Lamb_t\vv^{\text{min}}_{t}$,
where $\vv_{\max,t}$ and $\vv_{\min,t}$ are the singular row vector columns of $\V_t$ associated to the largest/smallest singular values of $\PP_t$.
Therefore, the product $\PPi_t = \A_t\PPi_{t-1}$ is given by
\begin{equation}
    \Pi_t = \left[\kappa_t\lambda^{\text{max}}_t\prod_{s=1}^{t-1}\left(\kappa_s\lambda^{\text{max}}_s\left(\uu^{\text{min}}_{s+1}\cdot\uu^{\text{max}}_{s}\right)\right)\right]\uu^{\text{max}}_{t}\left(\uu^{\text{min}}_1\right)^\dag + \cdots,
\end{equation}
so that $\pi_t$ \eqref{fhbgvqa} reads
\begin{equation}
    \ln\pi_t = \ln\left(\kappa_t\left|\lambda^{\text{max}}_t\right|\right) + \sum_{s=1}^{t-1}\ln\left[\kappa_s\left|\lambda^{\text{max}}_s\right|\left|\uu^{\text{min}}_{s+1}\cdot\uu^{\text{max}}_{s}\right|\right] + \cdots,
\end{equation}
By the CLT, $\ln\pi_t$ is distributed according to 
$\mathcal{N}(t\gamma_0, t\sigma_0^2)$ where 
the Lyapunov exponent is approximately given as
\begin{equation}
    \gamma_0 = \ln\lambda^{\text{max}} + \ln\kappa - \mu_u,
\end{equation}
Here, we define $\ln\lambda^{\text{max}} := \mathbb{E}\left[\ln\left|\lambda^{\text{max}}_t\right|\right]$ and
$\ln\kappa:=\mathbb{E}\left[\ln\kappa_t\right]$. The term $\mu_u := -\mathbb{E}\left[\ln\left|\uu^{\text{min}}_{t}\cdot\uu^{\text{max}}_{t-1}\right|\right]\ge 0$
is the average logarithmic penalty from angular misalignment and 
quantifies the systematic reduction of growth due to imperfect reinjection
into the most expanding direction after each random rotation.
Note that, when \( \mu_u = 0 \), i.e. when \( \uu^{\text{min}}_{t} = \uu^{\text{max}}_{t-1} \) almost surely, the reactive mode is perfectly reinjected into the non-normal direction at each step (see Fig.~\ref{fig:diagram}).
In this case, the dominant approximation of the Lyapunov exponent converges to the upper bound defined in \eqref{eq:lyapunov_bnd}.
Furthermore, the variance \( \sigma_0^2 \) consists of the contributions from the variances of the processes \( \ln \lambda^{\text{max}}_t \),
\( \ln \kappa_t \), and \( \ln \left| \uu^{\text{min}}_{t} \cdot \uu^{\text{max}}_{t-1} \right| \), as well as their covariances.
It also incorporate the effect of temporal autocorrelations in \( \ln \left| \uu^{\text{min}}_{t} \cdot \uu^{\text{max}}_{t-1} \right| \), leading to a tail exponent of the form
\begin{equation}
    \alpha \approx -2\frac{\gamma_0}{\sigma_0^2} .
\end{equation}

\vskip 0.2cm
\noindent
{\it Application to the Ginibre ensemble.}
Results from random matrix theory show that, for broad classes of ensembles,
eigenvectors and eigenvalues become asymptotically independent (see, e.g.,~\cite{TaoVu2012,Forrester2010}),
while perturbation arguments indicate that correlations between spectra and eigenbases are typically weak in high dimensions.  
In the Ginibre ensemble  \cite{ginibre1965statistical} (matrices with i.i.d. centered Gaussian entries, almost surely non-normal),
this asymptotic independence is rigorously established provided the system remains away from spectral criticality
(i.e., the spectral radius stays strictly inside the unit circle) and from eigenvalue degeneracies
(i.e., eigenvalues remain well separated)~\cite{bourgade2019distribution}.  
These two conditions ensure spectral stability, making the assumption of eigenvalue-eigenvector 
independence both natural and robust in the generic Ginibre case.

In high-dimensional random ensembles such as the Ginibre ensemble \cite{ginibre1965statistical}, 
the expected log-condition number scales as 
$\mathbb{E}[\ln \kappa] \sim \ln N$ \cite{chalker1998eigenvector}, 
so that non-normal amplification inevitably grows with system size. 
This makes non-normality the dominant route to instability and the emergence of power law statistics in large systems, providing a generic mechanism for scale-free behavior.

\vskip 0.2cm
\noindent
{\it Matrix Ensemble with Independent Spectral and Geometric Components.}
To illustrate our theoretical predictions, we perform numerical experiments based on a matrix ensemble $\A_t$ 
specifically designed to disentangle the effects of the spectrum, rotation, and shear induced by non-normality.
The setup mirrors the generic scenario outlined above while remaining analytically tractable. 
At each time step \( t \), the random update matrix is drawn according to 
\eqref{eq:apx_dec_n_dim23}, where \( \Lamb_t \), \( \U_t \), \( \V_t \), and \( \Sig_t \) 
are mutually independent across components at each step.
In light of known results for the Ginibre ensemble, this construction is consistent 
with sampling Ginibre matrices conditioned on having all eigenvalues inside the unit 
circle and sufficiently far from the unit root. 
For simplicity, we assume that \( \U_t \) and \( \V_t \) are Haar-distributed on the 
unitary (or orthogonal) group. Owing to the independence between spectral properties, 
shear, and rotational components, this framework allows for a sharper analytical 
characterization of the dominant expansion, leading to explicit approximations of 
the Lyapunov exponent and the associated tail exponent.

\emph{Condition number from Extreme Value Theory (EVT).} 
The condition number $\kappa_t = s_{\max,t}/s_{\min,t}$ is
defined as the ratio of the largest to smallest singular values of the eigenbasis transformation matrix $\PP_t$. 
Assuming that the logarithms of the singular values are i.i.d., 
$\ln s_{i,t} \sim \mathcal{N}(0,\sigma^2)$,
extreme value theory implies that, for large $N$,
\begin{equation}
\ln \kappa_t \;\simeq\; 2\sigma \sqrt{2\ln N}, 
\qquad
\sigma_\kappa^2 \;\simeq\; \frac{\pi^2\sigma^2}{6\ln N},
\label{eq:kappa_stats}
\end{equation}
where $\sigma_\kappa^2$ denotes the variance of $\ln \kappa_t$. 
Consequently, we expect the Lyapunov exponent to scale as
\begin{equation}
    \gamma \sim \sigma \sqrt{\ln N}.
\end{equation}
It is worth emphasizing that, within this ensemble, 
$\ln \kappa_t$ grows like $\sqrt{\ln N}$, 
whereas in the Ginibre ensemble it scales as $\ln N$. 
The price paid for disentangling non-normal shear, rotation, 
and spectral properties, thereby achieving analytical control, is thus a more conservative scaling of the degree of non-normality with system size.

\paragraph*{Near-critical tail exponent.}
Since the Lyapunov exponent $\gamma$ increases with the degree of non-normality, 
and since the two main control parameters are the variance $\sigma^2$ of the logarithm of the singular values 
and the system dimension $N$, 
it follows that, for a fixed distribution, there exists a critical dimension $N_c$ 
such that $\gamma \approx 0$ at $N = N_c$. 
At this threshold, the system approaches criticality and the tail exponent correspondingly decreases toward $\alpha \approx 0$.
More generally, the tail exponent decreases \emph{linearly} with the distance to criticality. 
The linear rate at which the tail exponent decreases as the system approaches criticality scales as $\sqrt{2\ln N}$, 
with a denominator collecting the variance contributions from the spectrum, the condition number, and the reinjection mechanism.
In high dimensions, the interplay among these three sources of randomness 
creates a subtle balance between stability and instability.
In particular, the critical non-normal variance $\sigma_c(N)$ scales as $\sqrt{\ln N}$, 
implying that larger systems can sustain greater heterogeneity before destabilizing 
(i.e., before $\gamma$ becomes positive). 
As $\sigma \to \sigma_c(N)$, the tail exponent $\alpha$ decreases continuously toward zero, 
indicating the emergence of increasingly heavy-tailed fluctuations 
even when the spectrum remains stable. 
Thus, dimensionality and non-normal amplification jointly push the system toward effective criticality, 
manifested by diverging moments and vanishing tail exponents.

\vskip 0.2cm
\noindent
{\it Numerical Illustration.}
For the numerical simulations, we sample the matrices $\A_t$ such that their eigenvalues satisfy
$\ln \lambda_i \sim \mathcal{N}(-1,\,0.1)$,
thereby preventing $\Lamb_t$ from being (approximately) proportional to the identity. 
This choice ensures that $\ln \rho \approx -1$, so that the system remains spectrally stable.
$\PP_t$ is sampled from its singular value decomposition (SVD),
$\PP_t = \U_t \Sig_t \V$,
where $\U_t$ and $\V$ are independent Haar-distributed unitary matrices,
and $\Sig_t$ is diagonal with i.i.d.\ singular values $s_{i,t}$ satisfying
$\ln s_{i,t} \sim \mathcal{N}(0,\sigma)$.
The role of $\sigma$ in shaping the dynamics will be examined.
The matrix $\V^\dagger \Lamb \V$ is drawn once at the beginning of the simulation and defines the purely ``normal'' component of the dynamics. 
The factors $\U_t \Sig_t$ then encode the non-normal stretching and rotational effects. 
By construction, these contributions modify the transient amplification properties 
without altering the spectrum of the matrix at any time.

We choose to sample the matrices $\PP_t$ in this manner, rather than directly from the Ginibre ensemble, 
because it provides explicit control over the statistics of the rotational components $\U_t$, $\V$, 
as well as over the condition number $\kappa_t$. 
In addition, this construction avoids the need to numerically invert $\PP_t$, 
an operation that would be prone to severe instability due to the high probability of ill-conditioning.
One consequence of this construction is that the expected value of the log condition number grows only as 
$\ln \kappa \sim \sqrt{\ln N}$ (\ref{eq:kappa_stats}) with matrix dimension, implying a slower scaling of $\kappa$ with $N$ 
than in the Ginibre ensemble. In the latter case, the typical condition number grows linearly with dimension, 
$\kappa \sim N$, leading to $\mathbb{E}[\ln \kappa] \sim \ln N$. 
As a result, the non-normal amplifications observed in our construction should be viewed as conservative estimates, 
since the generic Ginibre case would exhibit even stronger effects.
A further advantage is that this ensemble introduces two natural control parameters:  
(i) the \emph{volatility of the log singular values} ($\sigma$);  
(ii) the \emph{dimension of the system} ($N$).  
These two parameters can be combined into $\ln \kappa$, which, by Extreme Value Theory (EVT), scales as $\ln \kappa \sim \sigma \sqrt{\ln N}$.
This setup ensures spectral stability, 
while fluctuations in $\kappa_t$ probe non-normal amplification. 

\begin{figure}
    \centering
    \includegraphics[width=0.23\textwidth, trim=0 345 0 0 ,clip]{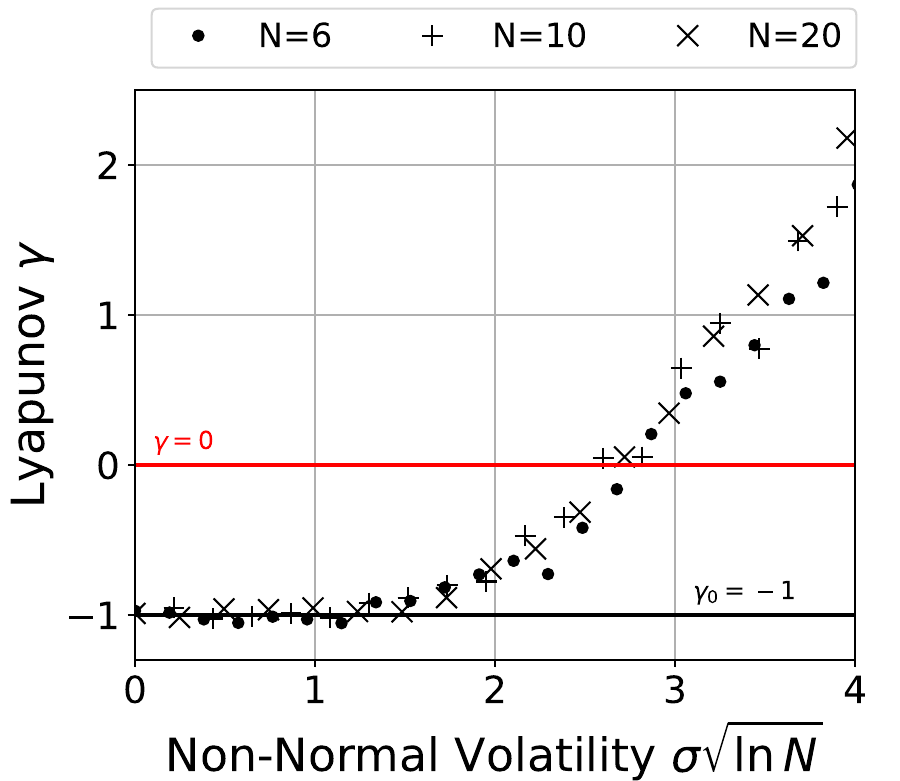}
    \includegraphics[width=0.23\textwidth, trim=0 0 0 375 ,clip]{lyapunov_nd_main.pdf}
    \caption{
        Empirical Lyapunov and tail exponents measured for a $N$ dimensions Kesten process (\ref{eq:apx_dec_n_dim23}) for $N=6,\,10,\,20$,
        following the numerical procedure defined in the text.
        The $x$-axis corresponds to the expected log-condition number 
        $\mathbb{E}[\ln \kappa] \sim \sigma \sqrt{\ln N}$ in the extreme-value limit,  
        with i.i.d. log-normal singular values $\ln s_{i,t} \sim \mathcal{N}(0,\sigma^2)$. 
        The red horizontal line marks the critical threshold $\gamma=0$. \\
        The numerical procedures used to estimate Lyapunov and tail exponents are detailed in the SM.
    }
    \label{fig:lyapunov}
\end{figure}

Figure~\ref{fig:lyapunov} illustrates how the Lyapunov exponent depends on the singular value dispersion $\sigma$ and the system size $N$.  
The main findings are: (i) for normal systems ($\sigma=0$), $\gamma$ is independent of $N$, as expected;  
(ii) introducing non-normality ($\sigma>0$) leads to an approximately linear growth of $\gamma$ with $\sigma$,
with a larger slope when random rotations are present;  
(iii) the dominant finite-$N$ effect is well captured by $\gamma \approx \gamma_0 + \mathbb{E}[\ln \kappa]$,
where $\mathbb{E}[\ln \kappa]\sim \sqrt{\ln N}$, consistent with extreme-value arguments.  
At small $\sigma$, the growth of $\gamma$ is initially quadratic,
before crossing over to the linear regime.
Overall, the simulations confirm that non-normality raises the Lyapunov exponent,
and the scaling of the log-condition number follows the theoretical $\sqrt{\ln N}$ behavior.

Figure~\ref{fig:lyapunov} also reports the behavior of the power law tail exponent $\alpha$.  
The main finding is the confirmation of the prediction that $\alpha$ decreases approximately linearly with $\sigma_c-\sigma\sqrt{\ln N}$ when $\sigma$ is close to but smaller than the critical value $\sigma_c$,
where the Lyapunov exponent $\gamma$ crosses zero.  
In particular, the left panel shows that $\gamma\approx 0$ for $\sigma\sqrt{\ln N}\approx 2.7$,
which is consistent with the linear extrapolation of $\alpha$ for $\alpha < 5$ observed in the right panel of Fig.~\ref{fig:lyapunov}.  
Moreover, $\alpha$ exhibits an approximate scaling with $\sigma\sqrt{\ln N}$,
vanishing as $\sigma\to\sigma_c(N)$, in agreement with the simultaneous vanishing of $\gamma$.  
The right panel of Fig.~\ref{fig:lyapunov} further reveals a crossover to an approximately constant $\alpha$ for $\sigma\sqrt{\ln N}<1.5$,
corresponding to the disappearance of non-normal eigenvector amplification.  
A particularly noteworthy outcome is that the theoretical predictions,
which are derived under asymptotic assumptions such as the law of large numbers,
the central limit theorem, and large-$N$ arguments, are already well supported by simulations with moderate system sizes $N=6$ to $20$.  
This robustness illustrates how asymptotic results can hold quantitatively even for relatively small $N$.

\noindent
{\it Polymer stretching in turbulent flows as an instance of eigenvector amplification}.
The coil-stretch transition of single molecules in turbulent or random flows is described by repeated action of random velocity-gradient matrices, with entropic elasticity providing the reinjection mechanism that prevents collapse and giving rise to a power law distribution of the end-to-end polymer lengths~\cite{Balkovsky2000,Chertkov2000}. 
This behavior can be formalized by starting from the stochastic evolution equation for the end-to-end extension vector $\mathbf{R}$ of the polymer molecule,
$\dot{\mathbf{R}} = (\nabla \mathbf{v})\,\mathbf{R} - \tfrac{1}{\tau}\mathbf{R}$,
where $\nabla \mathbf{v}$ is the local velocity-gradient tensor and $\tau$ the
polymer relaxation time.
After time discretization, the polymer dynamics reduces exactly to a multidimensional Kesten recursion driven by velocity-gradient matrices and additive thermal noise.,
where the evolution matrix is
$\A_t = \left(1 - \tfrac{\delta t}{\tau}\right)\I + \delta t \nabla \mathbf{v}_t$.
In turbulent and random flows, $\nabla \mathbf{v}_t$ is typically strongly non-normal,
with large eigenvector overlaps and pseudospectral amplification \cite{Trefethen1993,Farrell1988}.
As a result, the operator norm of $\A_t$
can exceed its spectral radius by orders of magnitude.
This places polymer stretching within the precise mathematical framework analyzed in this work.
Introducing the normalized orientation
$\mathbf{n}=\mathbf{R}/|\mathbf{R}|$, the instantaneous stretching rate is
reduced to the scalar process $z(t) = n_a n_b \,\partial_a v_b = \mathbf{n}^\top (\nabla \mathbf{v}) \,\mathbf{n}$
and the log-extension obeys $ r(t) = \ln\!\left(\tfrac{R(t)}{R_0}\right) = \int_0^t (z(t') - 1/\tau)\,dt'$.
Using large-deviation theory for the time-integrated process $z(t)$, Refs.\cite{Balkovsky2000,Chertkov2000} show
that $r$ acquires an exponential tail, which translates into a power law tail
for $R$: $P(R)\sim R^{2a-1}$.

Viewed through our framework, the origin of these heavy tails becomes transparent. 
The key variable $z(t)$ depends not only on the eigenvalues of $\nabla \mathbf{v}$ 
but also on its projection onto the instantaneous orientation $\mathbf{n}$. 
Because velocity-gradient tensors in turbulence are inherently non-normal, 
their eigenvectors are non-orthogonal; transient alignments between $\mathbf{n}$ 
and nearly co-linear eigendirections of $\nabla \mathbf{v}$ thus generate amplification 
events far larger than eigenvalues alone predict.
These rare constructive alignments constitute the large-deviation events governing 
the tail of $P(R)$, providing a concrete physical basis for the observed power law 
distribution of polymer extensions.
Within our framework, fluctuations of the eigenvector condition number $\kappa_t$ of the velocity-gradient tensor at each step increase the effective Lyapunov exponent by $\ln\kappa:=\mathbb{E}\left[\ln\kappa_t\right]$ and reduce the tail exponent.
Polymer stretching thus provides a concrete realization of the predicted shift $\alpha \to \alpha - {2 \ln\kappa \over \sigma_0^2}$.

\bibliography{bibliography} 

\section*{Numerical tools: Lyapunov \& Tail-Exponent}

\subsection{Numerical Estimation of Lyapunov Exponents}
\label{apx:lyapunov}

In this appendix, we briefly describe the numerical method used to estimate the Lyapunov exponents 
of products of random matrices, which is based on the classical \emph{QR reorthonormalization} 
technique introduced by Benettin et al.~\cite{Benettin1980,Benettin1980b}. 
This procedure is significantly more stable and accurate than a naive ``brute force'' approach.

\subsubsection{Brute Force Estimation.}

A straightforward way to approximate the largest Lyapunov exponent $\gamma$ is to iterate
\begin{equation}
    \PPi_t = \prod_{s=1}^t \A_s, \qquad
    \gamma \approx \frac{1}{t} \ln \|\PPi_t\|,
\end{equation}
for large $t$.  
However, this method is numerically unstable: the norm $\|\PPi_t\|$ grows (or decays) exponentially fast, 
leading to rapid loss of numerical precision.  
In addition, if one is interested in the \emph{full spectrum} of exponents (not just the largest), 
the brute force method is not applicable, since it only tracks the growth rate along a single direction.

\subsubsection{QR Reorthonormalization (Benettin Method)}

The Benettin method overcomes these issues by evolving an orthonormal basis of vectors instead of a single one.  
At each iteration, the random matrix $\A_t$ is applied to the current basis, 
and the resulting set of vectors is reorthonormalized via a QR decomposition:
\begin{equation}
    \A_t \Q_{t-1} = \Q_t \R_t,
\end{equation}
where $\Q_t$ is an orthogonal matrix and $\R_t$ is upper triangular.  
The diagonal elements of $\R_t$ encode the local stretching factors of the basis vectors.  
By accumulating the logarithms of these diagonal entries over time, 
we obtain estimates of the Lyapunov exponents:
\begin{equation}
    \gamma_i = \lim_{t \to \infty} \frac{1}{t} 
    \sum_{s=1}^t \ln |(R_s)_{ii}|,
    \qquad i = 1,\dots,N.
\end{equation}

This procedure has several advantages:
\begin{itemize}
    \item It maintains numerical stability, since the vectors are continuously reorthonormalized 
        and do not collapse into the dominant eigen-direction.
    \item It provides the \emph{entire Lyapunov spectrum}, not just the maximal exponent.
    \item Averaging over multiple independent realizations (ensembles) further reduces statistical noise 
        and yields estimates of standard errors.
\end{itemize}

\subsubsection{Practical Considerations}

In practice, the algorithm proceeds as follows:
\begin{enumerate}
    \item Initialize with an orthonormal basis $\Q_0 = I$.
    \item For each time step, multiply the basis by a random matrix $\A_t$ 
        and apply QR decomposition to reorthonormalize.
    \item Accumulate the logarithms of the diagonal entries of $\R_t$, 
        and normalize by the total time horizon $t_{\max}$ to estimate the exponents.
    \item Optionally, repeat over multiple ensembles to improve statistical accuracy 
        and compute error bars.
\end{enumerate}

This method is therefore more reliable than brute force multiplication, 
particularly in high-dimensional settings or when long time horizons are required.  
Throughout our numerical analysis, we employ this QR reorthonormalization scheme 
to compute the Lyapunov spectrum of Kesten processes, 
and we refer the reader to this appendix for methodological details.

\subsection{Numerical Estimation of the Tail Exponent}
\label{apx:tail}

Here, we describe the methodology used to estimate the tail exponent $\alpha$ 
of heavy-tailed distributions.  
Naively, one might attempt to fit a straight line in log--log coordinates to the survival function
\begin{equation}
    \mathbb{P}[X > x] \sim C x^{-\alpha},
\end{equation}
and take the slope as an estimate of $\alpha$.  
However, such linear regression on doubly logarithmic plots is statistically inefficient and strongly biased:
the choice of fitting range is arbitrary, correlations between points are ignored, 
and finite-sample fluctuations often dominate the visual slope.  
For these reasons, modern approaches rely instead on maximum-likelihood estimation (MLE) 
combined with goodness-of-fit diagnostics.  
In our work, we adopt the Clauset--Shalizi--Newman (CSN) method~\cite{Clauset2009}, 
which has become a standard in the analysis of heavy-tailed data.

\subsubsection{Pareto Maximum-Likelihood Estimator}

Suppose we fix a threshold $x_{\min}$ above which the data are assumed to follow a Pareto law.  
Given $N$ observations $\{x_i \geq x_{\min}\}$, the log-likelihood of a Pareto tail is maximized by
\begin{equation}
    \hat{\alpha}(x_{\min}) 
    = \frac{N}{\sum_{i=1}^N \ln\!\left(\frac{x_i}{x_{\min}}\right)}.
\end{equation}
This estimator is unbiased in the limit $N \to \infty$, and is statistically more efficient 
than slope-fitting in log--log space.

\subsubsection{Choosing the Threshold}

A central difficulty is the choice of $x_{\min}$: too low and the data deviate from a pure power law, 
too high and too few points remain in the tail.  
The CSN method selects $x_{\min}$ by minimizing the Kolmogorov--Smirnov (KS) distance 
between the empirical cumulative distribution function (CDF) of the tail 
and the fitted Pareto CDF.  
This balances goodness-of-fit with statistical power.

\subsubsection{Goodness-of-Fit and Uncertainty}

To assess whether the Pareto model is plausible, the CSN method uses a bootstrap procedure:  
synthetic datasets are generated from the fitted model and refitted in the same way, 
yielding a distribution of KS statistics.  
The $p$-value is the fraction of synthetic datasets with KS statistic at least as large as the empirical one.  
This provides a rigorous test of the hypothesis ``the tail follows a power law''.  
In addition, confidence intervals for $\alpha$ can be obtained from the bootstrap samples.

\subsubsection{Alternative Estimators}

Another widely used approach is the Hill estimator, which directly uses the top $k$ order statistics:  
\begin{equation}
    \hat{\alpha}_{\mathrm{Hill}}(k) 
    = \frac{k}{\sum_{i=1}^k \ln\!\left(\frac{x_{(i)}}{x_{(k)}}\right)},
\end{equation}
where $x_{(i)}$ denotes the $i$-th largest observation. 
One then analyzes the dependence of $\hat{\alpha}_{\mathrm{Hill}}(k)$ as a function of $k$. 
The presence of a plateau, when it exists, is typically interpreted as evidence of power law behavior, 
with the plateau value providing an estimate of the tail exponent. 
By contrast, the CSN method offers the advantage of automatically selecting the optimal range of $k$ 
and supplying associated statistical goodness-of-fit tests.

\subsubsection{Summary}

In summary, our estimation procedure consists of:
\begin{enumerate}
    \item Sorting the data and considering candidate thresholds $x_{\min}$.
    \item For each threshold, estimating $\alpha$ by MLE and computing the KS statistic.
    \item Selecting the $x_{\min}$ that minimizes the KS distance.
    \item Validating the fit and estimating confidence intervals via bootstrap resampling.
\end{enumerate}

This methodology avoids the pitfalls of linear regression in log--log space, 
yields statistically principled estimates of the tail exponent, 
and allows us to assess both parameter uncertainty and goodness-of-fit.  
In the numerical analysis, all reported values of $\alpha$ are obtained using this CSN framework.

\end{document}